\documentclass[multphys,natbib]{svmult}

\usepackage{graphicx}
\usepackage{makeidx}
\makeindex

\newcommand{\Msun}{\mathrm{M}_{\odot}}

\begin{document}

\title*{Formation of Globular Clusters in Hierarchical Cosmology: ART and Science}
\titlerunning{Formation of Globular Clusters}

\author{Oleg Y. Gnedin \and Jos\'e L. Prieto}
\authorrunning{Gnedin \& Prieto}

\institute{The Ohio State University,
           Department of Astronomy,
           Columbus, OH 43210, USA
\texttt{ognedin@astronomy.ohio-state.edu}}

\maketitle

{\bf Summary.}  We test the hypothesis that globular clusters form in
supergiant molecular clouds within high-redshift galaxies.  Numerical
simulations demonstrate that such large, dense, and cold gas clouds
assemble naturally in current hierarchical models of galaxy formation.
These clouds are enriched with heavy elements from earlier stars and
could produce star clusters in a similar way to nearby molecular
clouds.  The masses and sizes of the model clusters are in excellent
agreement with the observations of young massive clusters.  Do these
model clusters evolve into globular clusters that we see in our and
external galaxies?  In order to study their dynamical evolution, we
calculate the orbits of model clusters using the outputs of the
cosmological simulation of a Milky Way-sized galaxy.  We find that at
present the orbits are isotropic in the inner 50 kpc of the Galaxy and
preferentially radial at larger distances.  All clusters located
outside 10 kpc from the center formed in the now-disrupted satellite
galaxies.  The spatial distribution of model clusters is spheroidal,
with a power-law density profile consistent with observations.  The
combination of two-body scattering, tidal shocks, and stellar
evolution results in the evolution of the cluster mass function from
an initial power law to the observed log-normal distribution.
However, not all initial conditions and not all evolution scenarios
are consistent with the observed mass function.

\medskip\noindent 
Proceedings of {\it Globular Clusters -- Guides to Galaxies}, March
6-10, 2006, University of Concepci\'on, Chile, ed. T. Richtler et al.
(ESO/Springer)

\begin{figure}
\centering
\includegraphics[width=3.1in]{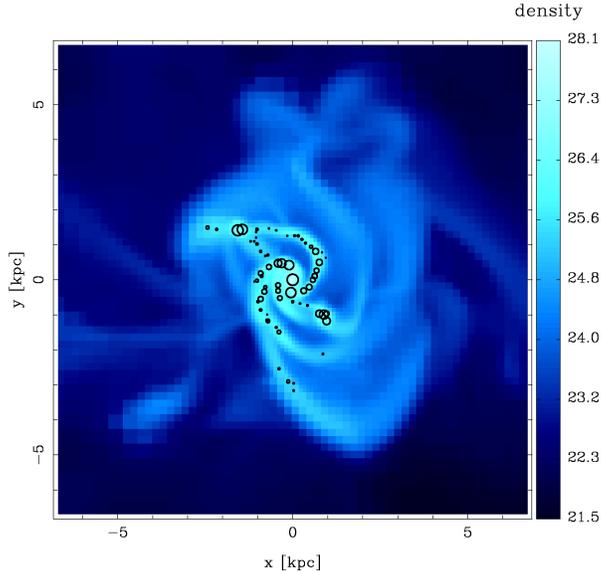}
\caption{A massive gaseous disk with prominent spiral arms, seen
  face-on at redshift $z=4$ in the process of active merging.  The gas
  density is projected over a $3.5$~kpc slice.  In our model star
  clusters form in giant gas clouds, shown by circles with the sizes
  corresponding to the cluster masses.  From
  \protect\citet{kravtsov_gnedin05}.}
  \label{fig:gc}
\end{figure}

\section{Giant Molecular Clouds at High Redshift}

The outcomes of many proposed models of globular cluster formation
depend largely on the assumed initial conditions.  The collapse of the
first cosmological $10^6\ \Msun$ gas clouds, or the fragmentation of
cold clouds in hot galactic corona gas, or the agglomeration of
pressurized clouds in mergers of spiral galaxies could all, in
principle, produce globular clusters, but only if those conditions
realized in nature.  Similarly, while observational evidence strongly
suggests that all stars and star clusters form in molecular clouds,
the initial conditions for cloud fragmentation are a major uncertainty
of star formation models.

The only information that we actually have about the initial
conditions comes from the early universe, when primordial density
fluctuations set the seeds for structure formation.  These
fluctuations are probed directly by the anisotropies of the cosmic
microwave background radiation.  Cosmological numerical simulations
study the growth of these fluctuations via gravitational instability,
in order to understand the formation of galaxies and all other
structures in the Universe.  The simulations begin with tiny
deviations from the Hubble flow, whose amplitudes are set by the
measured power spectrum of the primordial fluctuations while the
phases are assigned randomly.  Therefore, each particular simulation
provides only a statistical description of a representative part of
the Universe, although current models successfully reproduce major
features of the observed galaxies.

\citet{kravtsov_gnedin05} attempted to construct a first
self-consistent model of star cluster formation, using an
ultrahigh-resolution gasdynamics cosmological simulation with the
Adaptive Refinement Tree (ART) code.  They identified supergiant
molecular clouds in high-redshift galaxies as the likely formation
sites of globular clusters.  These clouds assemble during gas-rich
mergers of progenitor galaxies, when the available gas forms a thin,
cold, self-gravitating disk.  The disk develops strong spiral arms,
which further fragment into separate molecular clouds located along
the arms as beads on a string (see Fig. \ref{fig:gc}).

In this model, clusters form in relatively massive galaxies, with the
total mass $M_{\rm host} > 10^{9}\ \Msun$, beginning at redshift $z
\approx 10$.  The mass and density of the molecular clouds increase
with cosmic time, but the rate of galaxy mergers declines steadily.
Therefore, the cluster formation efficiency peaks at a certain
extended epoch, around $z \approx 4$, when the Universe is only 1.5
Gyr old.  The host galaxies are massive enough for their molecular
clouds to be shielded from the extragalactic UV radiation, so that
globular cluster formation is unaffected by the reionization of cosmic
hydrogen.  As a result of the mass-metallicity correlation of
progenitor galaxies, clusters forming at the same epoch but in
different-mass progenitors have different metallicities, ranging
between $10^{-3}$ and $10^{-1}$ solar.  The mass function of model
clusters is consistent with a power law $dN/dM \propto M^{-\alpha}$,
where $\alpha = 2.0 \pm 0.1$, similar to the observations of nearby
young star clusters.

\begin{figure}
\centering
\includegraphics[height=4.65in]{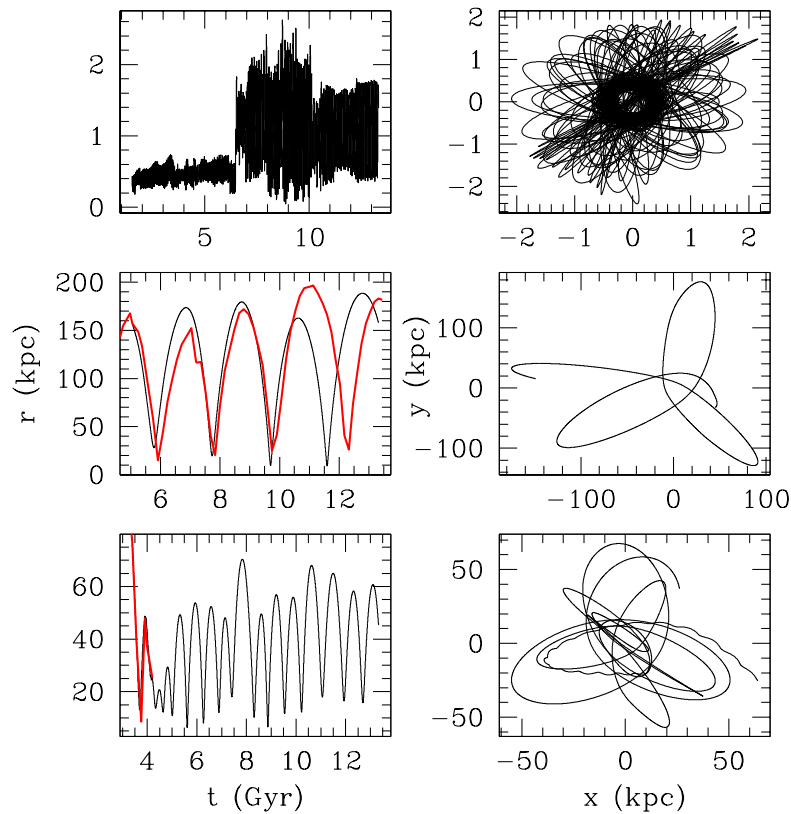}
\caption{Three types of globular clusters orbits.  Left panels show
  the distance to the center of the main halo, right panels show
  orbits in the plane of the main disk. {\em Top:} cluster formed in
  the main halo, on an initially circular orbit but was later
  scattered by accreted satellites.  {\em Middle:} cluster formed in a
  satellite halo, which survived as a distinct galaxy (thick red
  line).  {\em Bottom:} cluster formed in a satellite that was tidally
  disrupted at $t\simeq 4\ {\rm Gyr}$.}
  \label{fig:orbits}
\end{figure}

\begin{figure}
\centering
\includegraphics[height=2.25in]{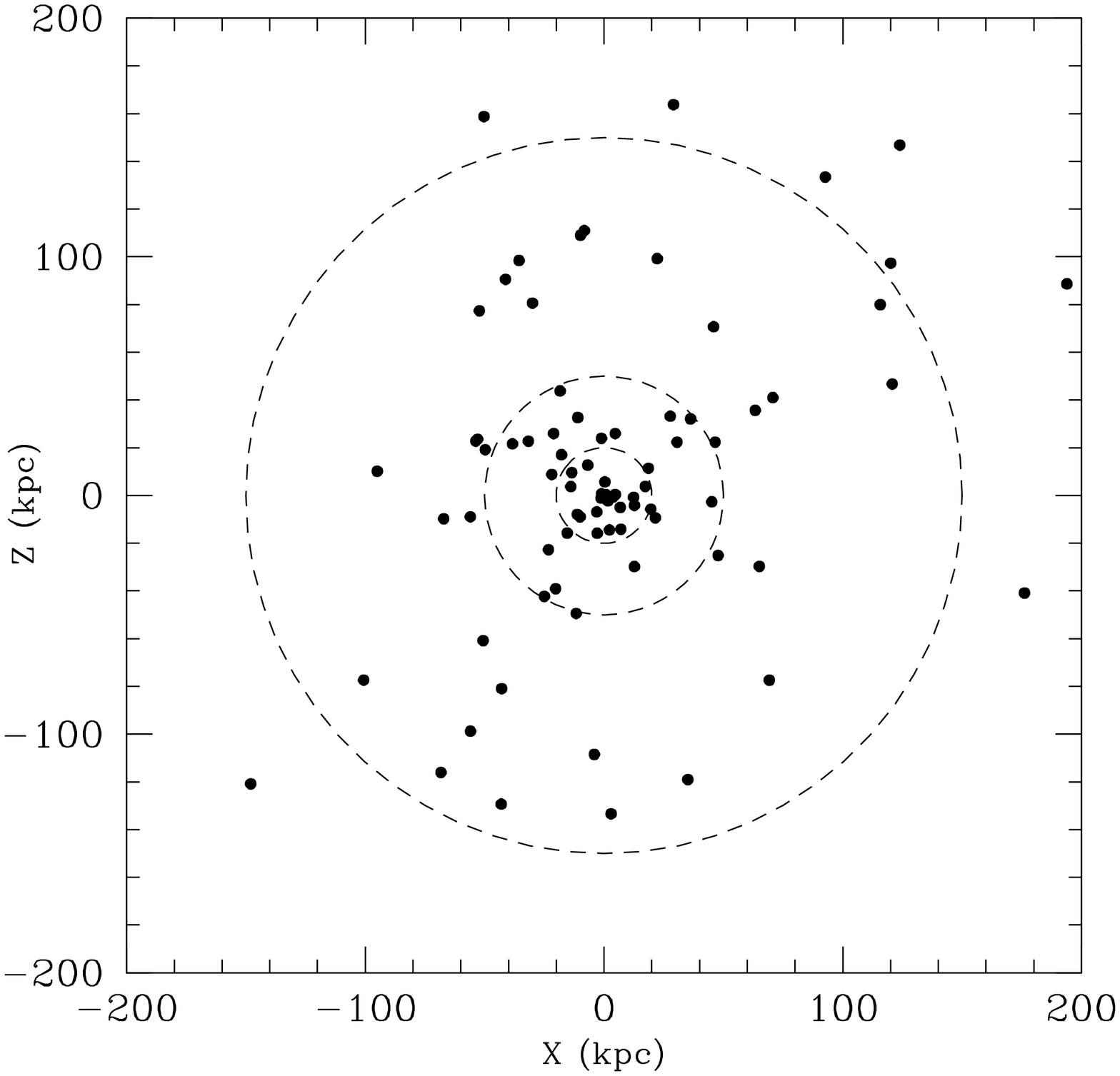}
\includegraphics[height=2.25in]{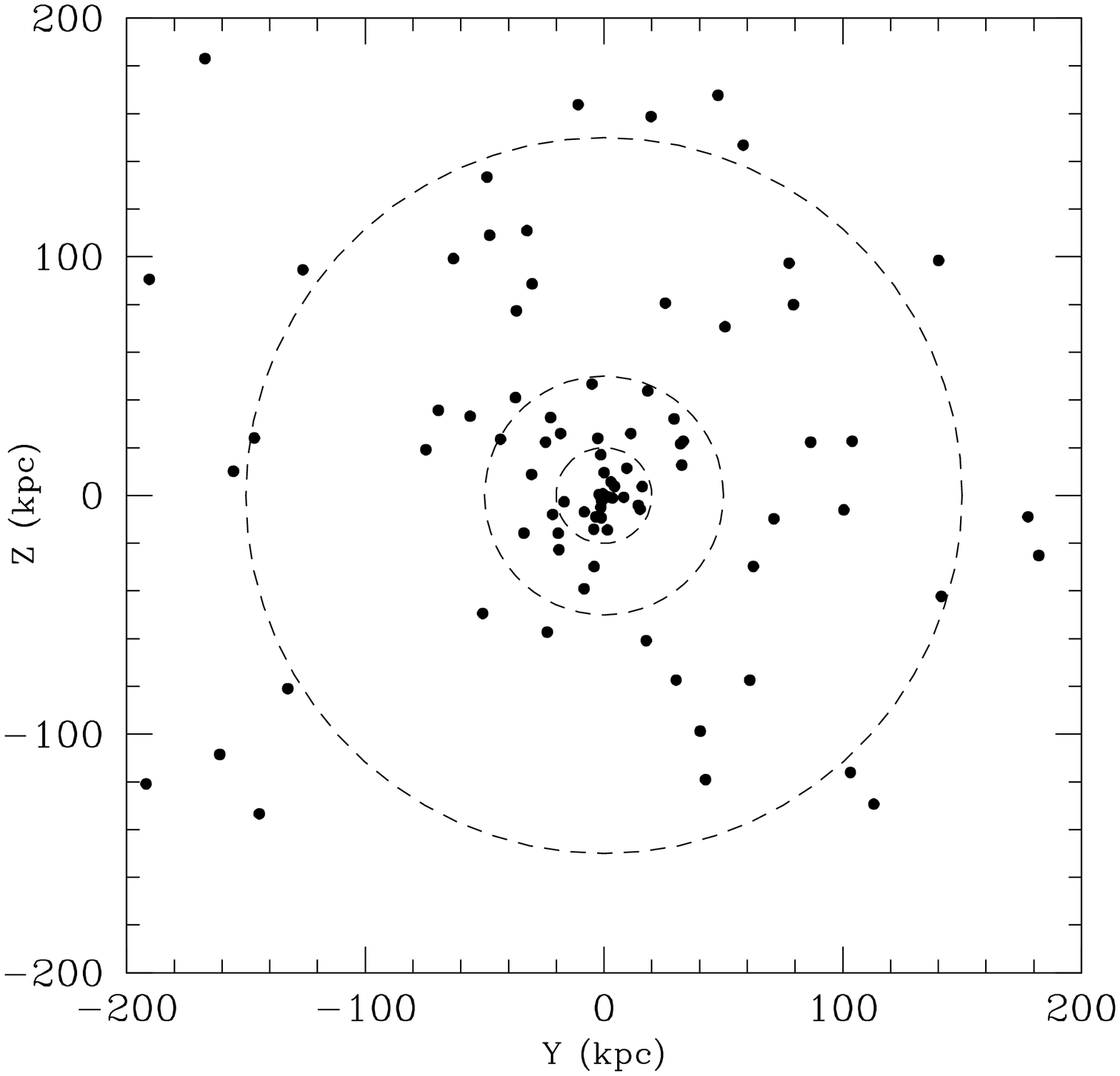}
\includegraphics[height=2.25in]{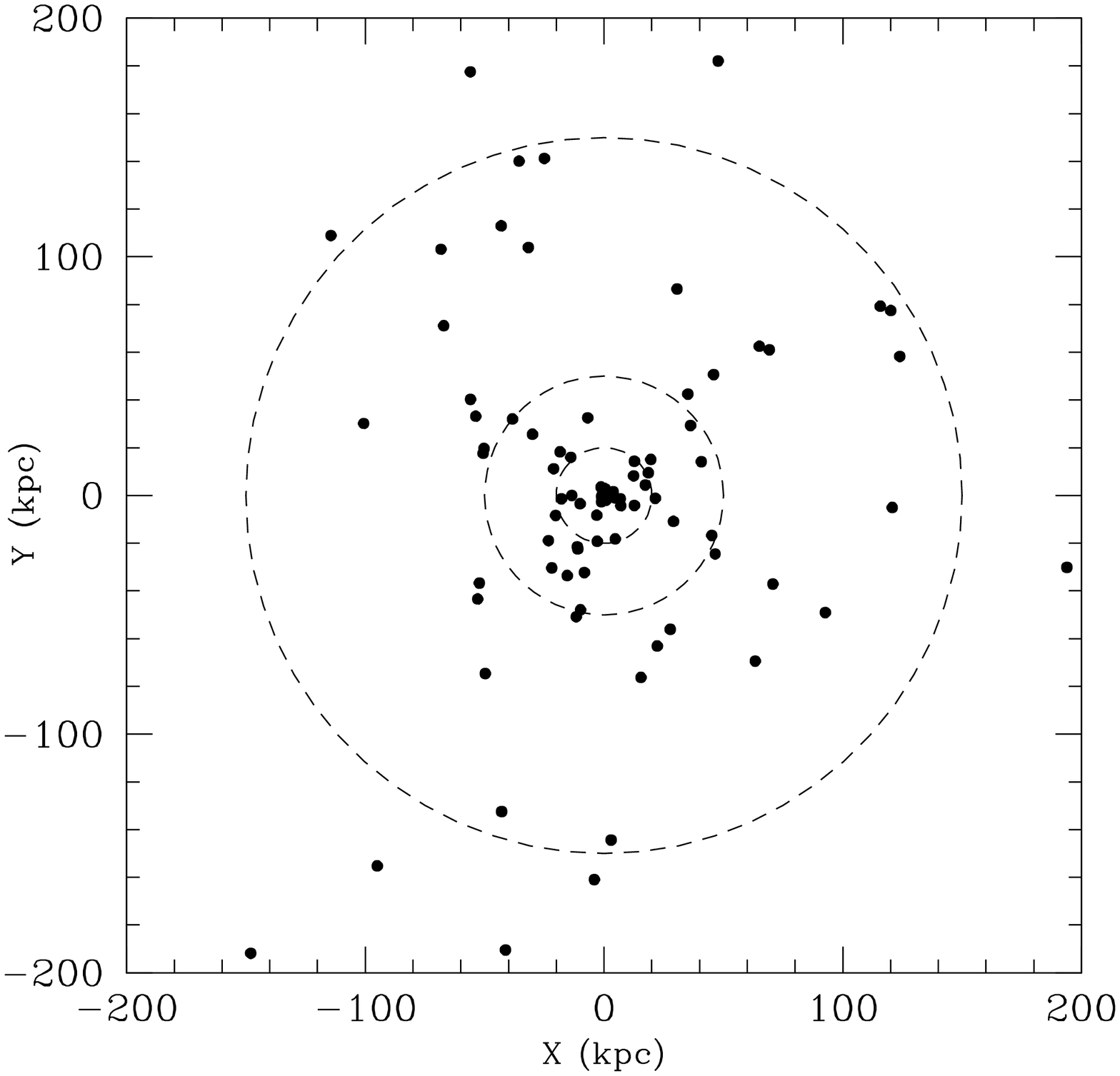}
\includegraphics[height=2.25in]{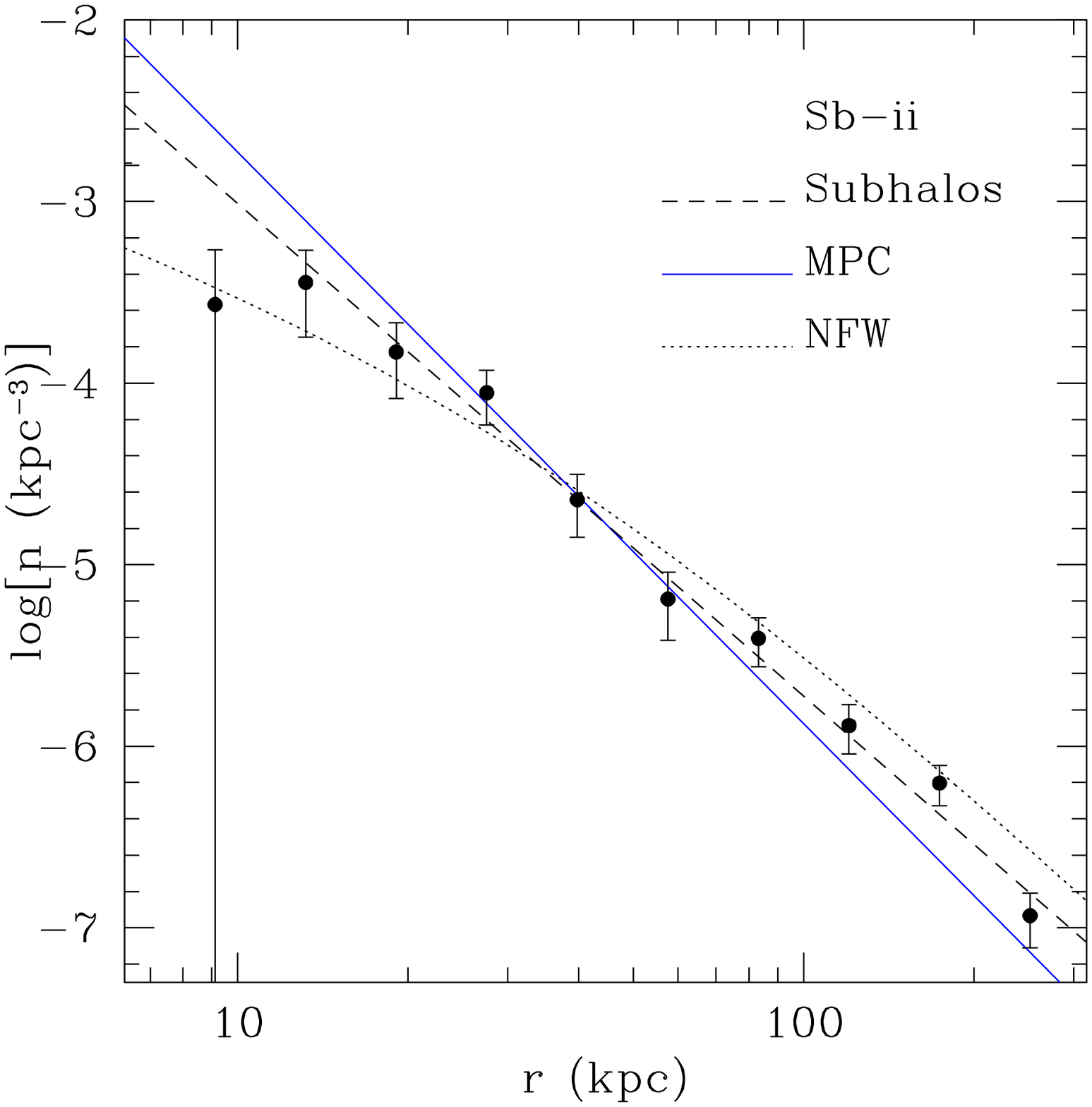}
\caption{Spatial distribution of surviving model clusters in the
  Galactic frame.  Dashed circles are at projected distances of 20,
  50, and 150 kpc.  The number density profile ({\it bottom right})
  can be fit by a power law, $n(r)\propto r^{-2.7}$.  The distribution
  of model clusters is similar to that of surviving satellite halos
  ({\it dashed line}) and smooth dark matter ({\it dotted line}).  It
  is also consistent with the observed distribution of metal-poor
  globular clusters in the Galaxy ({\it solid line}), plotted using
  the data from the catalog of \protect\citet{harris96}.}
  \label{fig:density}
\end{figure}

\begin{figure}
\centering \includegraphics[height=2.25in]{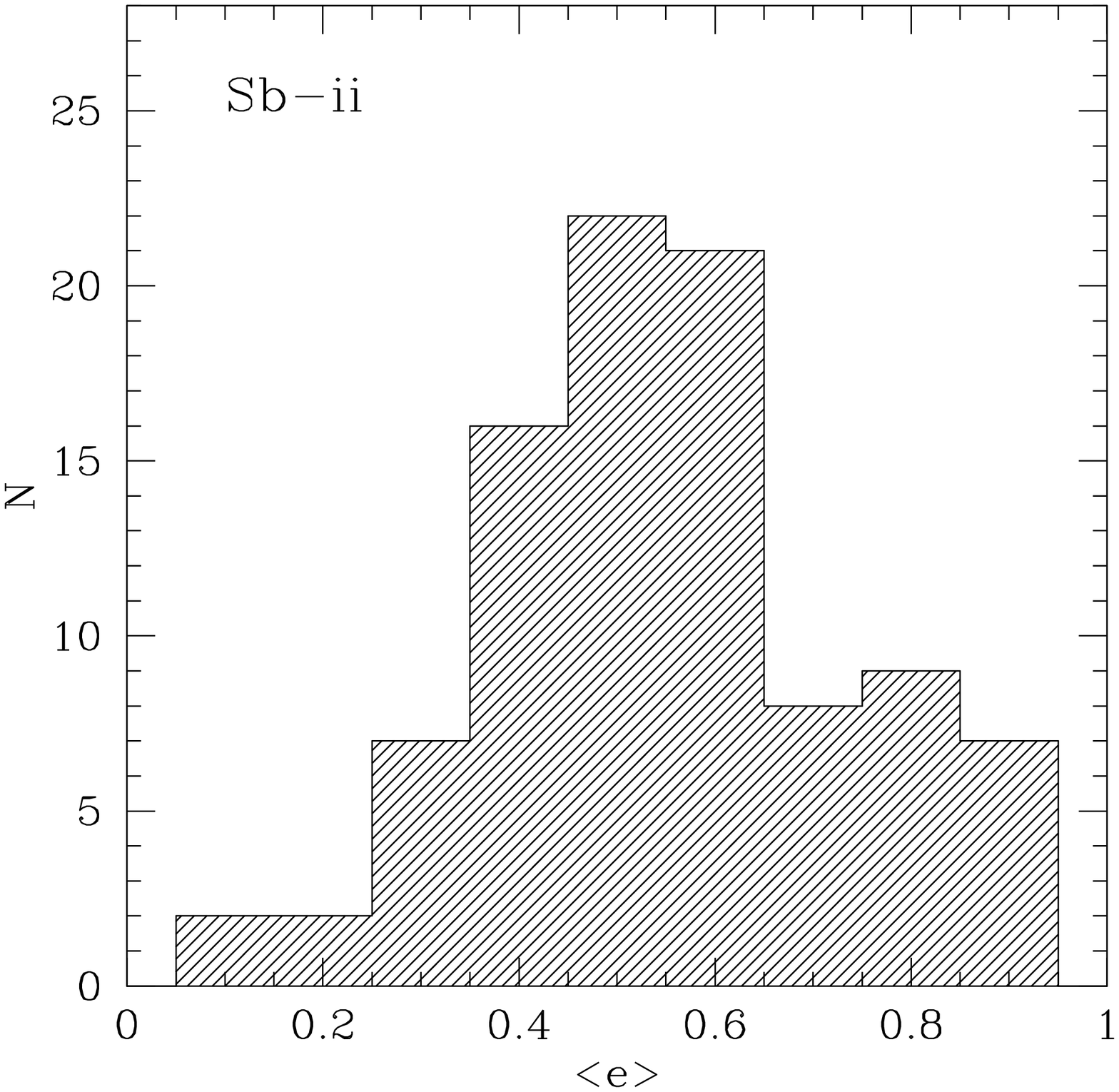}
\includegraphics[height=2.25in]{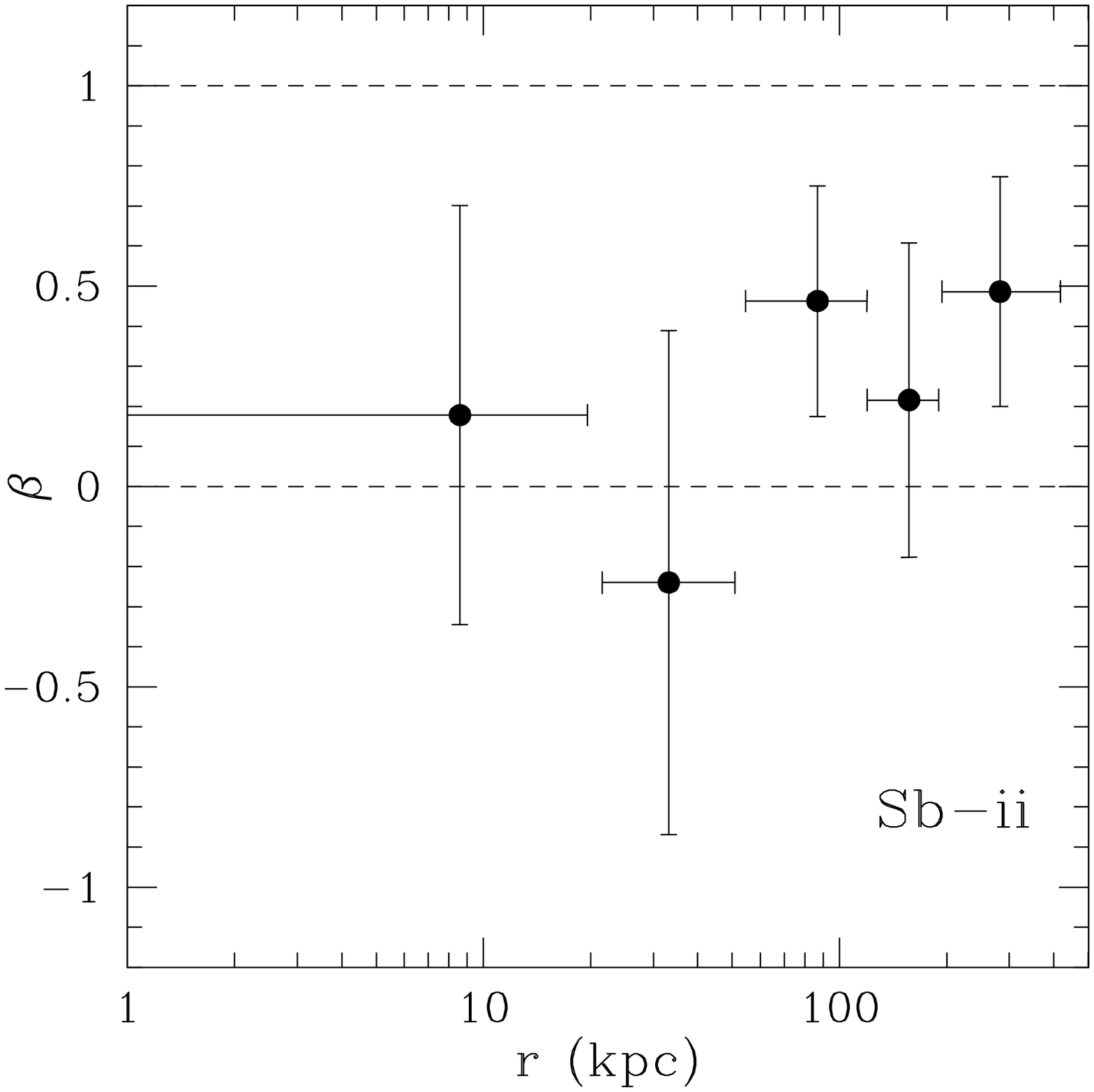}
\caption{{\it Left panel:} average eccentricity distribution of the
  surviving model clusters.  {\it Right panel:} Anisotropy parameter
  $\beta$ as a function of radius.  Vertical errorbars represent the
  error of the mean for each radial bin, while horizontal errorbars
  show the range of the bin.  Horizontal dashed lines illustrate an
  isotropic ($\beta=0$) and a purely radial ($\beta=1$) orbital
  distributions.}
  \label{fig:ecc}
\end{figure}

\section{Orbits of Globular Clusters}

We adopt this model to set up the initial positions, velocities, and
masses for our globular clusters.  We then calculate cluster orbits
using a separate collisionless $N$-body simulation described in
\citet{kravtsov_etal04}.  This is necessary because the original
gasdynamics simulation was stopped at $z \approx 3.3$, due to limited
computational resources.  By using the $N$-body simulation of a
similar galactic system, but complete to $z=0$, we are able to follow
the full dynamical evolution of globular clusters until the present
epoch.  We use the evolving properties of all progenitor halos, from
the outputs with a time resolution of $\sim 10^8$ yr, to derive the
gravitational potential in the whole computational volume at all
epochs.  We convert a fraction of the dark matter mass into the
analytical flattened disks, in order to model the effect of baryon
cooling and star formation on the galactic potential.  We calculate
the orbits of globular clusters in this potential from the time when
their host galaxies accrete onto the main (most massive) galaxy.
Using these orbits, we calculate the dynamical evolution of model
clusters, including the effects of stellar mass loss, two-body
relaxation, tidal truncation, and tidal shocks.

We consider several possible scenarios, one with all clusters forming
in a short interval of time around redshift $z=4$, and the others with
a continuous formation of clusters between $z=9$ and $z=3$.  Below we
discuss the spatial and kinematic distributions of globular clusters
for the best-fit model with the synchronous formation at $z=4$.

In our model, all clusters form on nearly circular orbits within the
disks of progenitor galaxies.  Present globular clusters in the Galaxy
could either have formed in the main disk, have come from the
now-disrupted progenitor galaxies, or have remained attached to a
satellite galaxy.  Figure \ref{fig:orbits} shows the three
corresponding types of cluster orbits.  Even the clusters formed
within the inner 10 kpc of the main Galactic disk do not stay on
circular orbits.  They are scattered to eccentric orbits by accreted
satellites, while the growth of the disk increases the average orbital
radius.  Triaxiality of the dark halo (not included in present
calculations) would also scatter the cluster orbits.  The clusters
left over from the disrupted progenitor galaxies typically lie at
larger distances, between 20 and 60 kpc, and belong to the inner halo
class.  Their orbits are inclined with respect to the Galactic disk
and are fairly isotropic.  The clusters still associated with the
surviving satellite galaxies are located in the outer halo, beyond 100
kpc from the Galactic center.  Note that these clusters may still be
scattered away from their hosts during close encounters with other
satellites and consequently appear isolated.

Mergers of progenitor galaxies ensure the present spheroidal
distribution of the globular cluster system (Fig. \ref{fig:density}).
Most clusters are now within 50 kpc from the center, but some are
located as far as 200 kpc.  The azimuthally-averaged space density of
globular clusters is consistent with a power law, $n(r)\propto
r^{-\gamma}$, with the slope $\gamma \approx 2.7$.  Since all of the
distant clusters originate in progenitor galaxies and share similar
orbits with their hosts, the distribution of the clusters is almost
identical to that of the surviving satellite halos.  This power law is
similar to the observed distribution of the metal-poor ($\mbox{[Fe/H]}
< -0.8$) globular clusters in the Galaxy.  Such comparison is
appropriate, for our model of cluster formation at high redshift
currently includes only low metallicity clusters ($\mbox{[Fe/H]} \le
-1$).  Thus the formation of globular clusters in progenitor galaxies
with subsequent merging is fully consistent with the observed spatial
distribution of the Galactic metal-poor globulars.

Figure \ref{fig:ecc} shows the kinematics of model clusters.  Most
orbits have moderate average eccentricity, $0.4 < \left< e \right> <
0.7$, expected for an isotropic distribution.  The anisotropy
parameter, $\beta = 1 - v_t^2/2v_r^2$, is indeed close to zero in the
inner 50 kpc from the Galactic center.  At larger distances, cluster
orbits tend to be more radial.  There, in the outer halo, host
galaxies have had only a few passages through the Galaxy or even fall
in for the first time.

\begin{figure}
\centering \includegraphics[height=2.65in]{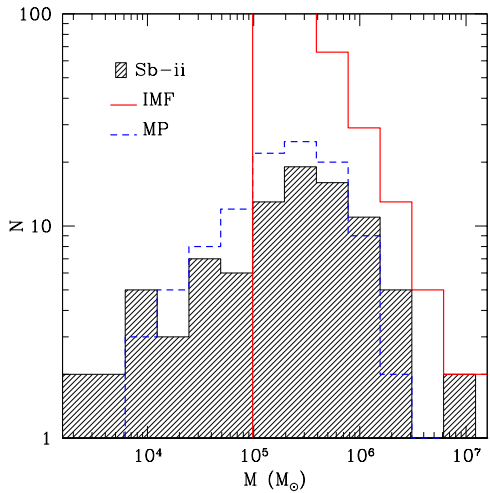}
\caption{Evolution of the mass function of model clusters from an
  initial power law ({\it solid line}) to a peaked distribution at
  present ({\it histogram}), including mass loss due to stellar
  evolution, two-body relaxation, and tidal shocks.  For comparison,
  dashed histogram shows the mass function of metal-poor globular
  clusters in the Galaxy.}
  \label{fig:mf}
\end{figure}

\section{Evolution of the Globular Cluster Mass Function}

Using these realistic orbits, we can now calculate the cluster
disruption rates.  Sophisticated models of the dynamical evolution of
globular clusters have been developed using direct $N$-body
simulations as well as the orbit-averaged Fokker-Planck and Monte
Carlo models.  They are described and referenced in many good reviews,
including \citet{spitzer87, gnedin_ostriker97, gnedin_etal99,
fall_zhang01, baumgardt_makino03}.  Several processes combine and
reinforce each other in removing stars from globular clusters: stellar
mass loss, two-body scattering, external tidal shocks, and dynamical
friction of cluster orbits.  The last three are sensitive to the
external tidal field and therefore, to cluster orbits.  While a
general framework for all these processes has been worked out already,
the knowledge of realistic cluster orbits is essential for accurate
calculations of the disruption.

Figure \ref{fig:mf} shows the transformation of the cluster mass
function from an initial power law, $dN/dM \propto M^{-2}$, into a
final bell-shape distribution.  In this model all globular clusters
form at the same redshift, $z=4$, or about 12 Gyr ago.  The half-mass
radii, $R_h$, are set by the condition that the median density,
$M/R_h^3$, is initially the same for all clusters and remains constant
as a function of time.  Over the course of their evolution, numerous
low-mass clusters are disrupted by two-body relaxation while the
high-mass clusters are truncated by tidal shocks.  The present mass
function is in excellent agreement with the observed mass function of
the Galactic metal-poor clusters.

\begin{figure}
\centering
\includegraphics[height=2.25in]{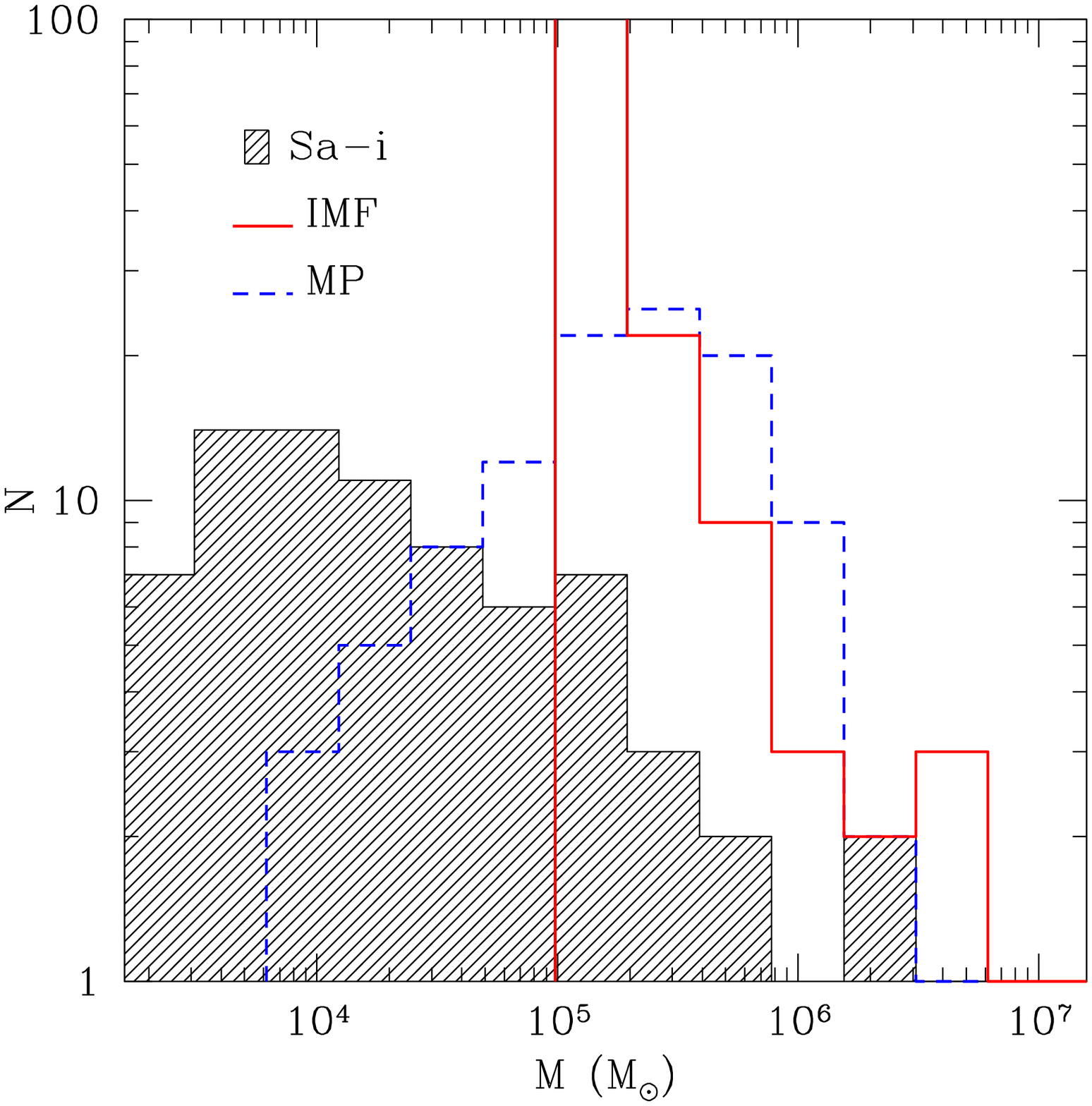}
\includegraphics[height=2.25in]{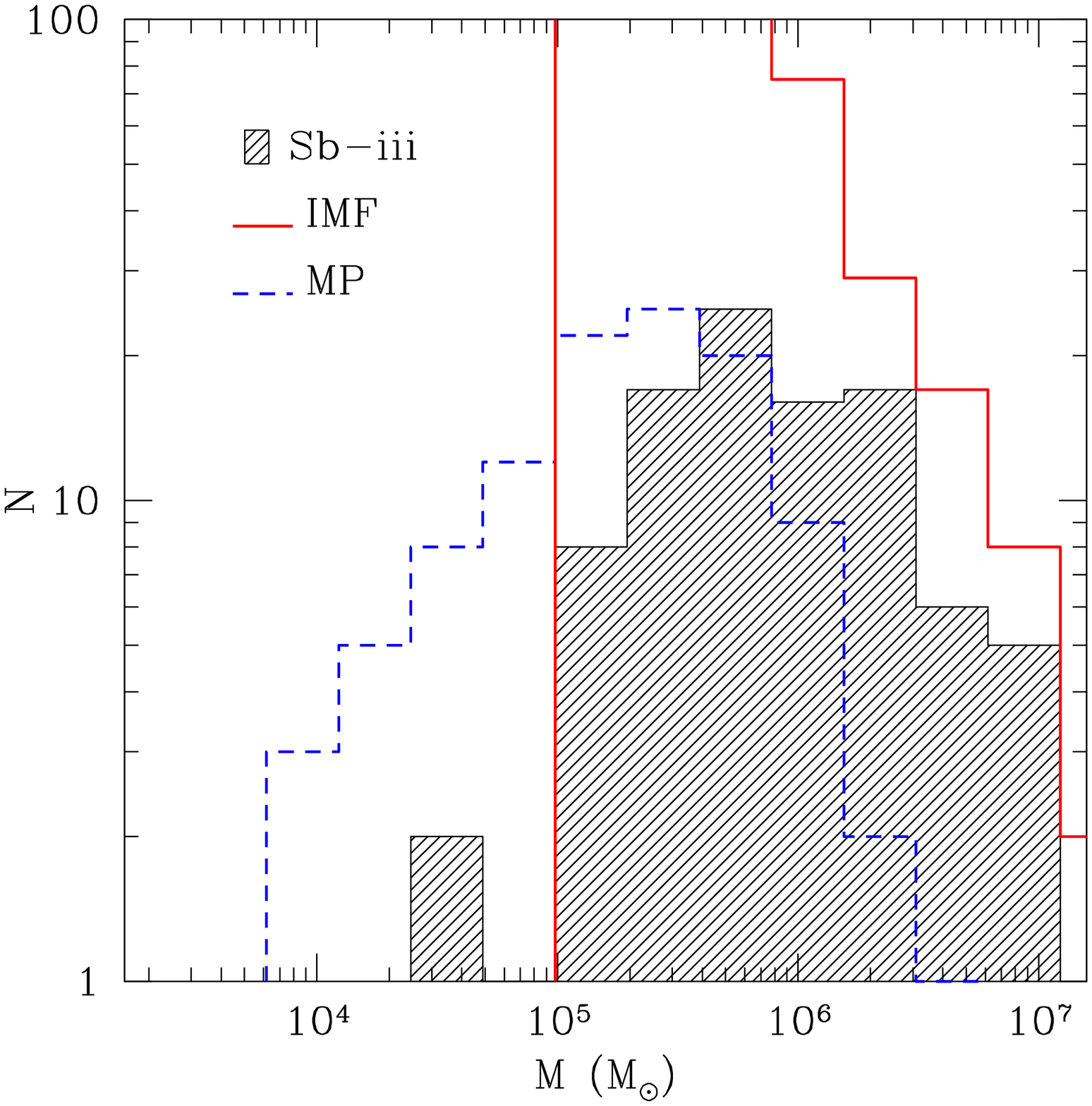}
\caption{Models that fail to reproduce the observed mass function of
  metal-poor globular clusters: with $R_h(t) = {\rm const}$ ({\it left})
  and with $R_h(t) \propto M(t)$ ({\it right}).}
  \label{fig:mf_fail}
\end{figure}

This result by itself is not new.  Previous studies of the evolution
of the cluster mass function have found that almost any initial
function can be turned into a peaked distribution by the combination
of two-body relaxation and tidal shocks.  However, the efficiency of
these processes depends on the cluster mass and size, $M(t)$ and
$R_h(t)$.  {\it The new result is that we find that not all initial
conditions and not all evolutionary scenarios are consistent with the
observed mass function.}

Figure \ref{fig:mf_fail} provides two examples.  In the first, the
half-mass radius is kept fixed at $R_h = 2.4$ pc (median value for
Galactic globulars) for clusters of all masses and at all times.  The
median density $M(t)/R_h^3$ thus decreases as the clusters lose mass.
Two-body scattering becomes less efficient and spares many low-mass
clusters, while tidal shocks become more efficient and disrupt most
high-mass clusters.  The final distribution is severely skewed
towards small clusters.

In the second example, the median density is initially fixed, as in
our main model, but the size is assumed to evolve in proportion to
the mass, $R_h(t) \propto M(t)$.  In this case the cluster density
increases with time.  As a result, all of the low-mass clusters are
disrupted by the enhanced two-body relaxation, while the high-mass
clusters are unaffected by the weakened tidal shocks.  The final
distribution is skewed towards massive clusters.

Only our best-fit model (Figs. \ref{fig:orbits} --\ref{fig:mf})
successfully reproduces the observed mass function and spatial
distribution of metal-poor globular clusters in Galaxy.  In future
work we will investigate the predicted properties of metal-rich
globular clusters and their dependence on galaxy formation history.

\bigskip\noindent 
OYG acknowledges the support of the American Astronomical Society and
the National Science Foundation in the form of an International Travel
Grant.

\bibliography{gc}

\printindex
\end{document}